\documentclass[twocolumn,10pt,showpacs,preprintnumbers,amsmath,amssymb]{revtex4}
\usepackage{graphicx}
\usepackage{dcolumn}
\usepackage{bm}
\def\bea {\begin{eqnarray}}
\def\eea {\end{eqnarray}}

\def\be {\begin{equation}}
\def\ee {\end{equation}}

\begin{document}
\title{Thermalization in small system of hadron gas \\
and high-multiplicity $pp$ events}
\author{\it Nachiketa Sarkar}
\author{\it Premomoy Ghosh}
\email{prem@vecc.gov.in}
\address {Variable Energy Cyclotron Centre, HBNI, 1/AF Bidhan Nagar, Kolkata 700 064, India}   
\begin{abstract}
We study the system-size dependence of Knudsen number, a measure of degree of thermalization, for hadron resonance gas that follows the Lattice-QCD equation of 
state at zero chemical potential. A comparison between Knudsen numbers for the $AuAu$ collisions at RHIC and the hadron gas of size similar to the size of high-multiplicity 
$pp$ events at LHC, reassures the applicability of hydrodynamics in interpreting the features of particle production in high-multiplicity $pp$ events.\\
\end{abstract}
\maketitle
\section{Introduction}
\label{}
The Quark-Gluon Plasma (QGP), a thermalized partonic matter predicted in quantum chromodynamics (QCD), got the experimental endorsement \cite {ref01, ref02, ref03, 
ref04} from the ultra-relativistic $AuAu$ collisions at the RHIC, where the $pp$ collisions served the role of the base-line in finding the signals for the collective medium. 
The experiments at the LHC have raised the significance of the study of multiparticle production in $pp$ collisions by extracting long-range two-particle angular 
correlations and obtaining the characteristic flow parameters in high-multiplicity events of $pp$ collisions at $\sqrt s$ = 7 and 13 TeV \cite {ref05, ref06, ref07}. \\

In spite of the strong experimental signatures, the collectivity in $pp$ collisions is disputed on the issue of thermalization in a small system of short lifetime that may 
be formed in the $pp$ events. At this stage, it is pertinent to assess the degree of thermalization of the system formed in high-multiplicity $pp$ events and compare it with 
that of the medium formed in relativistic heavy-ion collisions where the local thermodynamic equilibrium is undisputedly established through satisfactory descriptions of 
data by relativistic hydrodynamics \cite{ref08}. \\

The study of the bulk thermodynamic properties of the strongly interacting matter in the QCD framework has been made possible by the formulation of the Lattice QCD 
(LQCD) \cite{ref09}. Recent LQCD simulations at finite temperature reveal existence of a de-confined partonic phase at high temperature and a confined hadronic phase 
at low temperature. The hadronic phase of the QCD medium is successfully addressed also by the Hadron Resonance Gas (HRG) Model \cite {ref10, ref11, ref12}, in 
statistical thermodynamic framework. An ideal HRG, formulated with discrete mass spectrum of identified hadrons and resonances, has been reinforced with the ab-initio 
confirmation \cite {ref13, ref14, ref15, ref16} from the LQCD. Also, as yet unmeasured higher-mass hadron states, provided by the exponentially growing continuum mass 
spectrum, proposed \cite {ref17} by R. Hagedorn, have important contribution to the equation of state (EoS) \cite {ref18, ref19, ref20}  below the critical temperature ($T_{c}$) 
for the QCD change of phase.\\ 

In this article, to address the cardinal question on the degree of thermalization in small system of high-multiplicity $pp$ events, we consider HRG with the Hagedorn mass 
spectrum. The excluded volume effect \cite {ref21, ref22} is implemented in the HRG by considering hard core radius of the constituents of the system. Finally, we study 
the system-size dependence of the Knudsen number for finite-size hadron gas, complying with the LQCD EoS, to access the degree of thermalization in hadron gas that 
may be formed in the final stage of high-multiplicity $pp$ events.\\

\section{Hadron Resonance Gas Model}
\label{}
The partition function of $i^{th}$ particle in a grand canonical ensemble of ideal HRG can be written as \cite{ref11}:
\begin{equation}
\ln Z_{i}\textsuperscript{id}=\pm\frac{V g_i}{2\pi^2 }\int_{0}  ^\infty p^2 dp \ln \left\{1\pm \exp[-(E_i-\mu_i)/T] \right\}   
\end{equation}
where $E_{i}=\sqrt{p^2+m^2_{i}}$ and $\mu_{i} = B_i\mu_B + S_i\mu_s + Q_i\mu_Q $. The symbols carry their usual meaning. The $(+)$ and $(-)$ sign corresponds to 
fermions and bosons respectively. The pressure $P(T)$, energy density $\epsilon (T)$ and the number density $n(T)$ for ideal hadron resonance gas, at 
$\mu$ = 0, can be written as :
\begin{equation}
P^{id}(T) =  \pm \sum_i \frac{g_iT}{2\pi^2 } \int_{0}  ^\infty p^2 dp \ln \left\{1\pm \exp(-E_i/T) \right\}
\end{equation}
\begin{equation}
\epsilon^{id}(T) = \sum_i \frac{g_i}{2\pi^2 }\int_{0}  ^\infty \frac {p^2 dp} {\exp(E_i/T) \pm 1 }E_i 
\end{equation}
\begin{equation}n^{id}(T) = \sum_i \frac{g_i}{2\pi^2 }\int_{0}  ^\infty \frac {p^2 dp}  { \exp(E_i/T) \pm 1}
\end{equation}
The Hagedorn mass spectrum is given by \cite{ref23}:
\begin{equation}
\rho(m) = C \frac{\theta(m-M_0)}{(m^2+m^2_0)^a}\exp(\frac{m}{T_H})
\end{equation}
where $T_{H}$ is the Hagedorn temperature that determines the slope of the exponential part of the mass spectrum.
The effect of repulsive interactions at the short distances, particularly significant for density related observables, are incorporated in a thermodynamically consistent manner, 
through the Van der Waals Excluded Volume (EV) method \cite {ref22}. The volume of the system is substituted with an effective volume obtained by excluding the sum of 
volume, $v$ = 16$\pi r^3$/3 of the constituent hadrons of hard core radius, $r$. The inclusion of the Hagedorn states in the system of hadrons and resonances naturally 
leads one to adopt the Boltzmann approximation. The thermodynamic variables for such a system with the excluded volume effect can be written as \cite {ref19}:
\begin{eqnarray}
P^{H}_{EV}(T) ={{\kappa}} P^{H}(T)\\
\epsilon^{H}_{EV}(T) = \frac{\kappa{\epsilon^{H}(T)}}{1 + {v}{\kappa}{n}^{H}(T)}\\
n^{H}_{EV}(T) = \frac{\kappa{n^{H}(T)}}{1 + {v}{\kappa}{n}^{H}(T)}
\label{eq:n_H_EV}
\end{eqnarray}
where ${\kappa}$ (\textless 1) is the excluded volume suppression factor, given by $\kappa = \exp(-{{v}{p^{H}}} / {T})$ and
\begin{widetext}
\begin{eqnarray}
P^{H}(T) =\frac{T}{2\pi^2 }\int  dm   \int_{0}  ^\infty p^2 dp \exp\Big (-\frac{\sqrt{m^2+p^2}}{T}\Big)\Big[\sum_i g_{i}\delta(m-m_{i})+\rho(m)\Big]\\
\epsilon^{H}(T) = \frac{1}{2\pi^2 }\int dm \int_{0}  ^\infty {p^2dp}\sqrt{m^2+p^2}\exp \Big (-\frac{\sqrt{m^2+p^2}}{T}\Big)\Big[\sum_i g_{i}\delta(m-m_{i})+\rho(m)\Big]\\
n^{H}(T) = \frac{1}{2\pi^2 }\int  dm \int_{0}  ^\infty {p^2dp}\exp\Big (-\frac{\sqrt{m^2+p^2}}{T}\Big)\Big[\sum_i g_{i}\delta(m-m_{i})+\rho(m)\Big]
\end{eqnarray}
\end{widetext}

\section{Results}
\label{}
\begin{figure*}[htb!]
\centering
\includegraphics[scale=0.65]{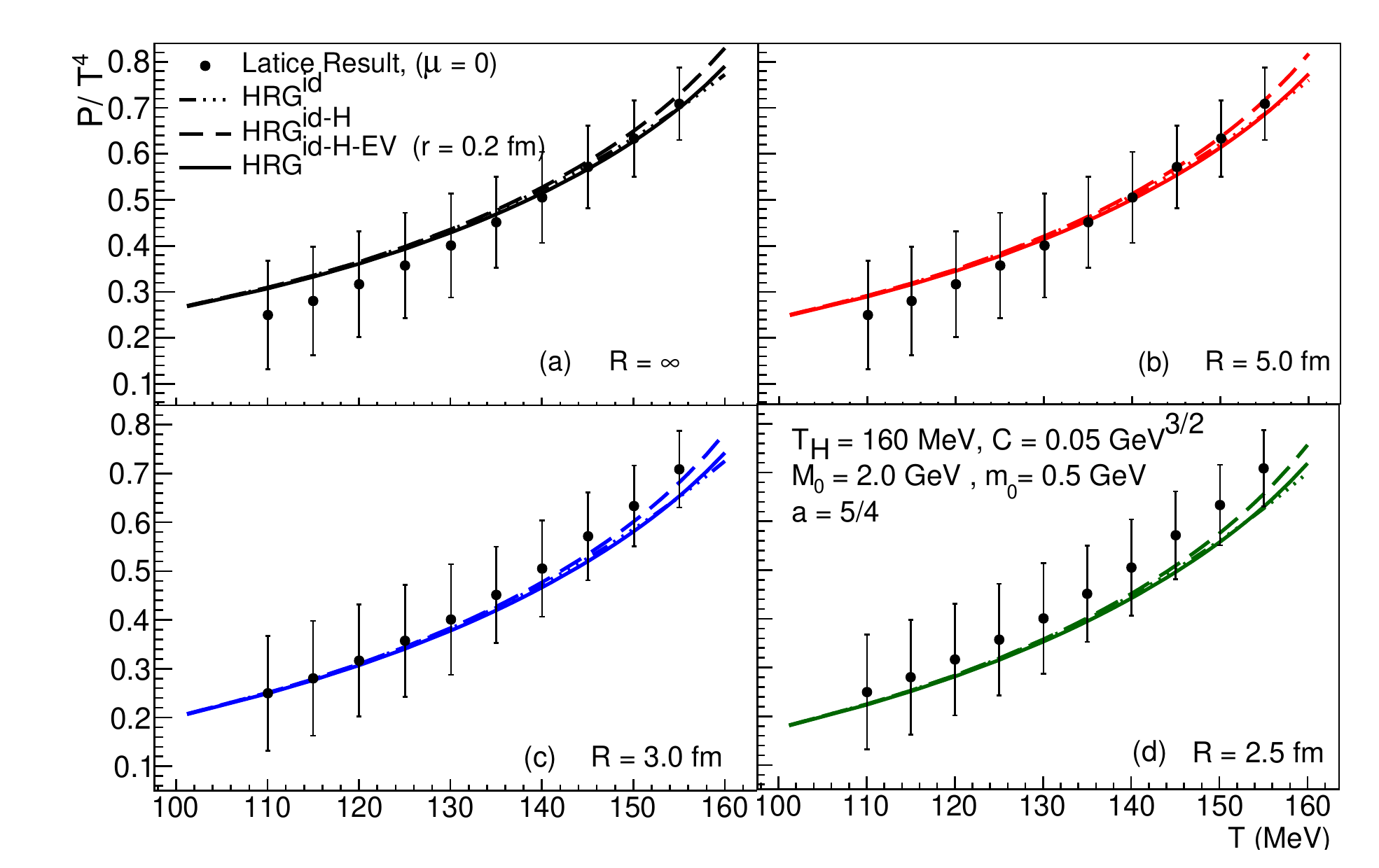}
\caption{The pressure for different system sizes: (a) infinite (b) $R$ = 5 fm, (c) $R$ = 3 fm and (d) $R$ = 2.5 fm of hadron gas with different effects as a function of $T$ is 
compared with LQCD data \cite {ref17}.}
\label{fig:pressure} 
\end{figure*}

\begin{figure*}[htb!]
\centering
\includegraphics[scale=0.65]{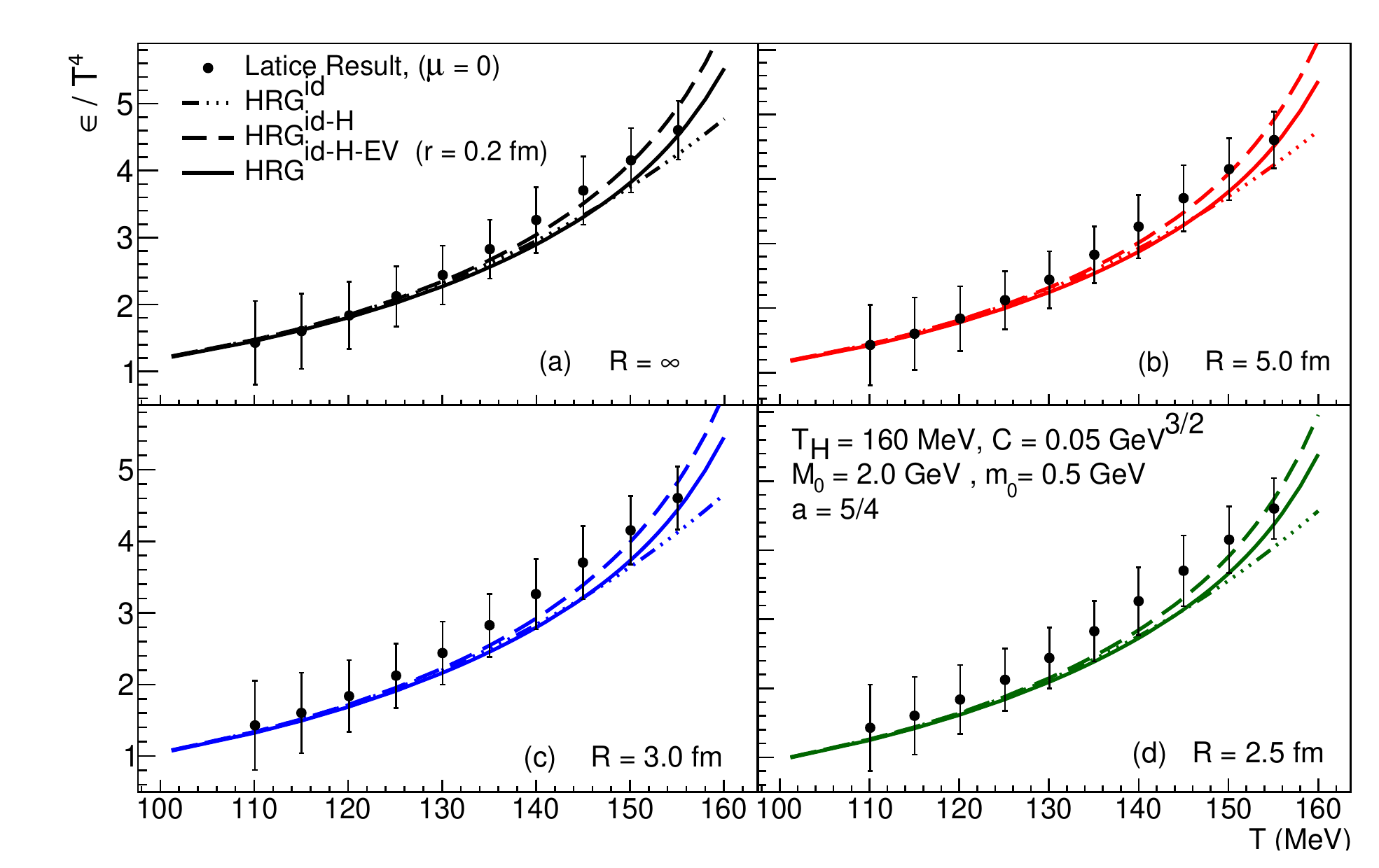}
\caption{The energy density for different options as mentioned in the caption of Figure~\ref{fig:pressure}.}
\label{fig:energy}  
\end{figure*}

To implement the finite size effect in hadron resonance gas, we start with infinite (in thermodynamic sense) system-size of hadron gas, incorporated with the Hagedorn 
states and excluded volume effect, where the LQCD EoS is contrasted satisfactorily. We optimize the values of the corresponding parameters, from the ranges as suggested 
in Ref. \cite {ref19}, for the simultaneous consideration of Hagedorn states and the excluded volume effect. The finite size effect on volume and number of particles, 
considered infinite in the thermodynamic limit, can be implemented \cite {ref24, ref25} by cutting off the low momentum regions in the integral over momentum space. We 
introduce the finite size effect \cite {ref25}, using the lower limit of momentum, $p_{cutoff} (MeV)$ = 197 ${\pi} / {{R} (fm)}$, where $R$ is the characteristic system-size. We 
consider the mass table of the PDG in Ref. \cite {ref26} and compare our calculations with the LQCD results of Ref. \cite {ref16}. The temperature dependence of pressure 
and energy-density of hadron gas for infinite as well as for a few representative finite sizes, $R$ = 2.5, 3 and 5 fm, for different options, 1) Ideal HRG, 2) Ideal HRG + 
Hagedorn States and 3) Ideal HRG + Hagedorn States + EV effect are presented in Figures~\ref{fig:pressure} and ~\ref{fig:energy}, respectively. It is clear from the figures 
that a system of hadrons, resonances and Hagedorn states up to the finite system-size of $R$ = 2.5 fm (corresponding to the $p_{cutoff} \approx$ 250 MeV), can be 
described by the LQCD EoS at zero chemical potential. \\

\begin{figure*}[htb!]
\centering
\includegraphics[scale=0.55]{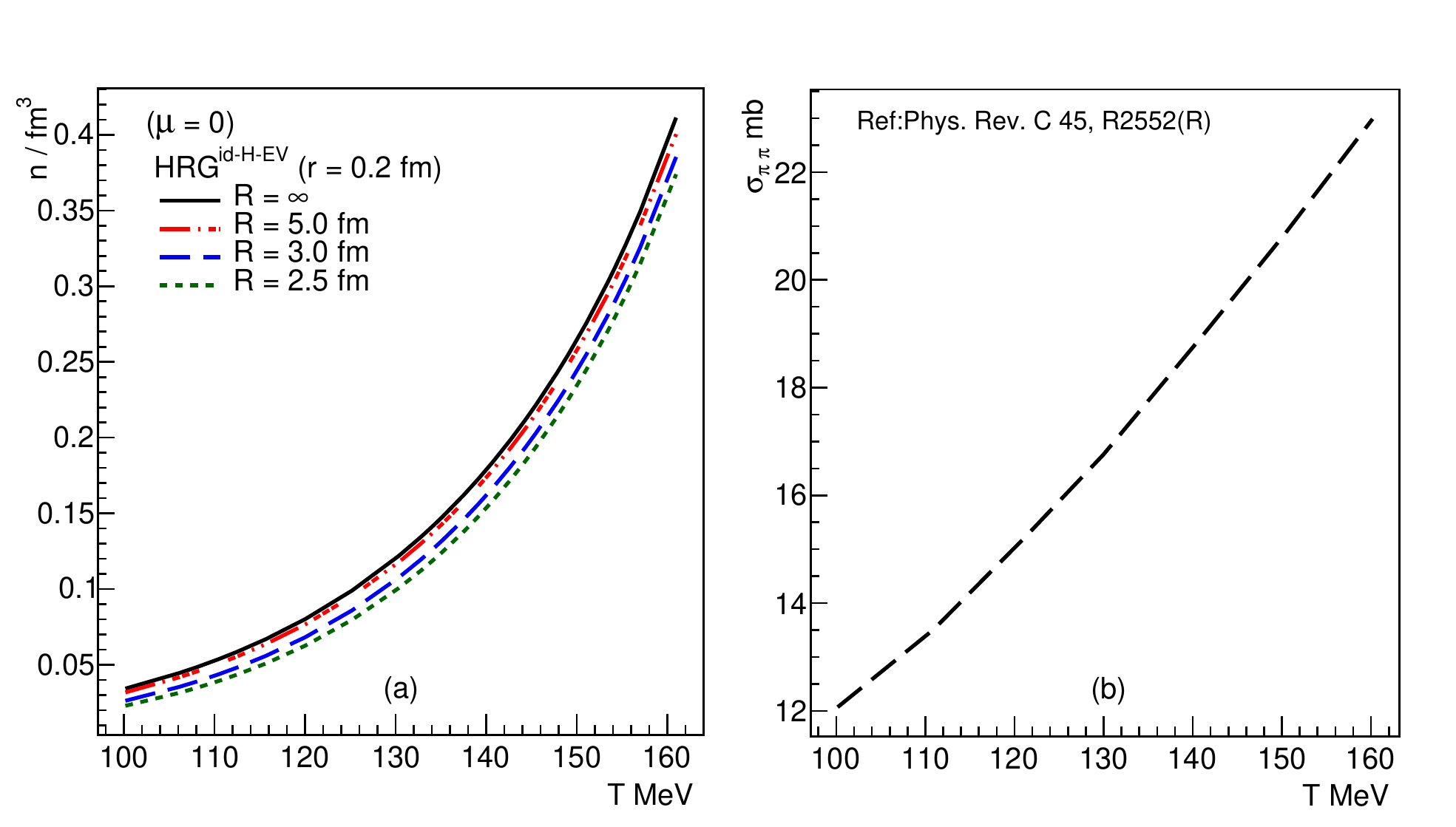}
\caption{(a)The number density for infinite as well as finite system sizes ($R$ = 5, 3 and 2.5 fm) of hadron resonance gas, including the Hagedorn 
States and the EV effect, as a function of temperature. (b) Corresponding $\sigma_{\pi \pi}$ in a thermalized pion gas as given in \cite {ref27}.}
\label{fig:number} 
\end{figure*}

\begin{figure}[htb!]
\centering
\includegraphics[scale=0.45]{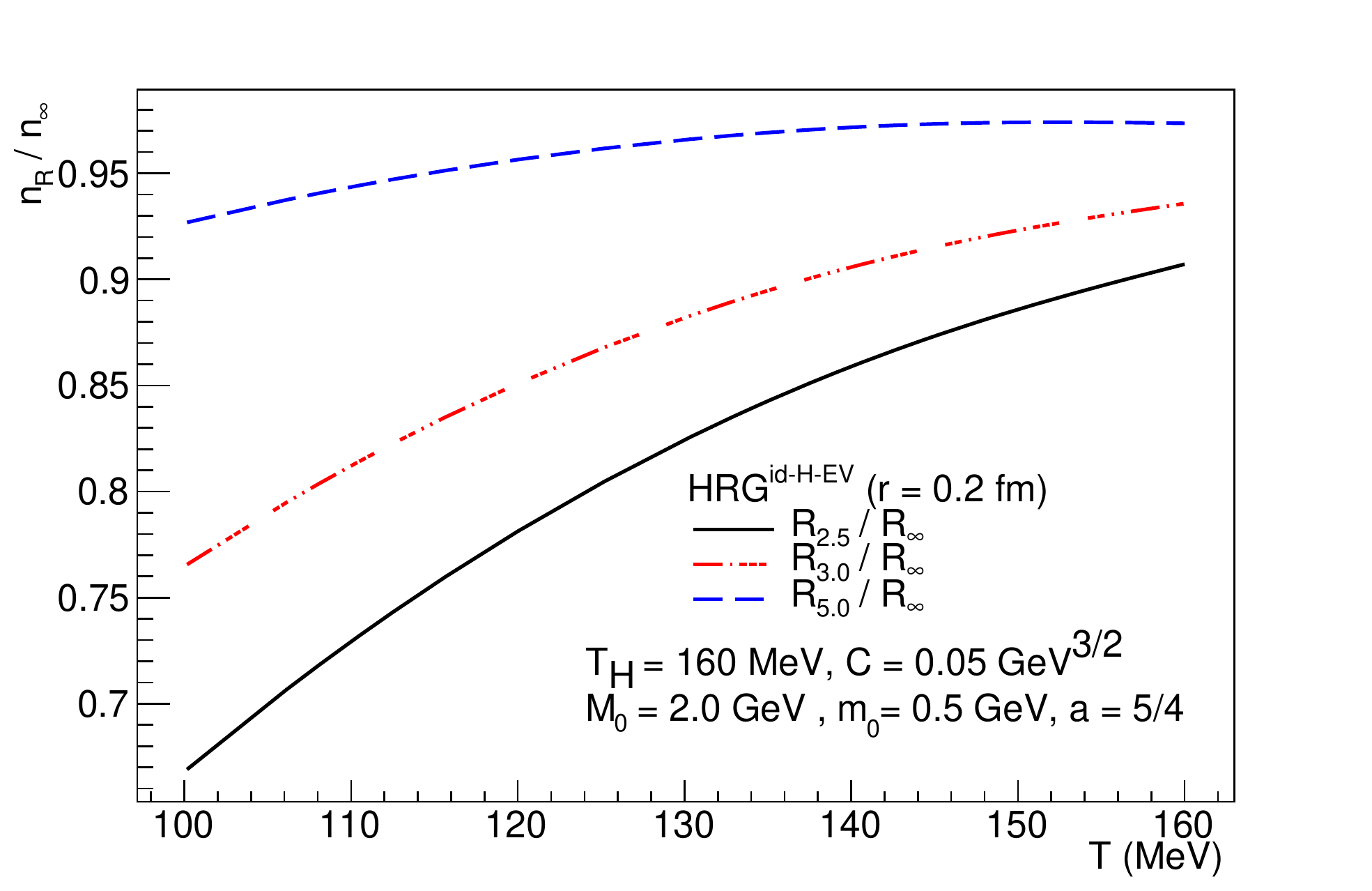}
\caption{The ratio of number density of finite ($R$ = 5, 3 and 2.5 fm) and the infinite system-size of hadron resonance gas, including the Hagedorn 
States and the EV effect as a function of temperature.}
\label{fig:ratio} 
\end{figure}

\begin{figure}[htb!]
\centering
\includegraphics[scale=0.45]{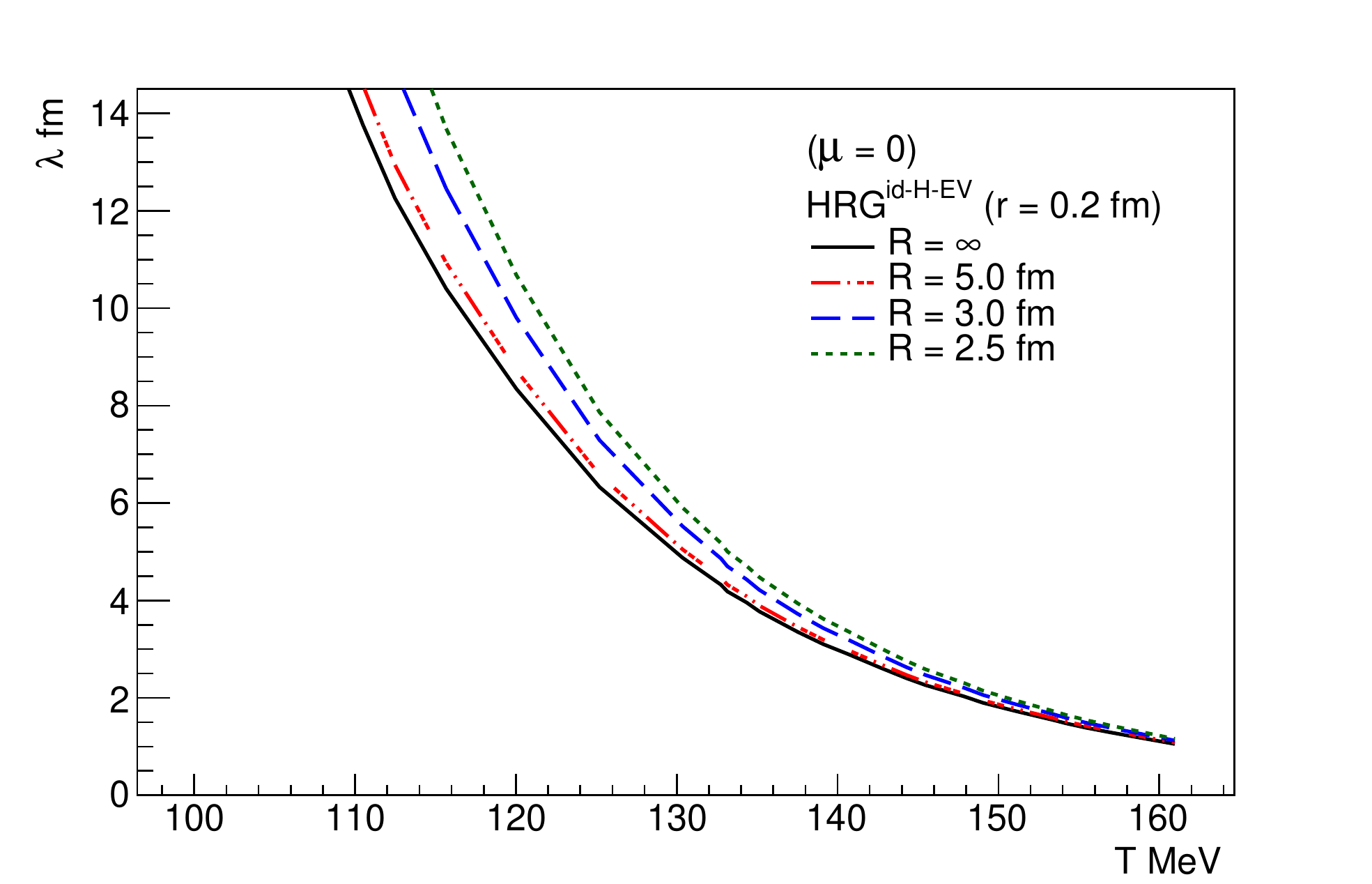}
\caption{Temperature dependent pion mean free path in different sizes of hadron gas, obtained by using the $\sigma_{\pi\pi}$ in a thermalized pion gas given in \cite {ref27}.}
\label{fig:freepath} 
\end{figure}

To calculate the mean free path, $\lambda = 1 / {n}{\sigma}$ of the constituents in the thermalized gas of hadrons, resonances and Hagedorn states, one needs to rely on 
approximations only, as the cross-sections for all the involved interactions are not available. Considering the fact that the pions are the most abundant constituents in an 
equilibrated hadron gas, for the present study, we approximate the temperature dependent number density (${n^{H}_{EV}(T)}$) given by equation~\ref{eq:n_H_EV} and 
depicted in the Figure~\ref{fig:number}(a) as representing the pion-density and compare the temperature dependent mean free path of pions in thermalized hadron gas of 
different system-size. The temperature dependent pion-pion cross-sections in thermalized pion gas, used for the temperature dependent mean free path calculations, has 
been obtained from the Ref.\cite {ref27} and presented in Figure~\ref{fig:number}(b). Figure~\ref{fig:number} shows that the number density varies with the system-size of 
the hadron gas all through the considered temperature range. Further study in terms of the ratio of number density of finite ($R$ = 5, 3 and 2.5 fm) and the infinite 
system-size of hadron resonance gas, reveals ( Figure~\ref{fig:ratio}) that the variation in the number density with the system-size decreases with increasing temperature. 
This feature of the number density for hadron gas of different sizes approaching closer value at high temperature is consistent with the observation \cite {ref25} on the 
temperature dependence of thermodynamical variables for the similar finite system-sizes of hadron gas. As has been pointed out in the Ref.~\cite {ref25}, the variation 
with the system-size at high temperature disappears due to the dominant population of higher mass resonances, including the Hagedorn States, which, unlike 
the low mass resonances, are not affected on implementation of finite size effect. As can be seen in the Figure~\ref{fig:freepath}, though the mean free path of pions for the 
hadron gas of different sizes, constrained with the LQCD EoS, is different in the lower region of the temperature scale, it asymptotically approaches the same value of $\sim$ 
1 fm at $T \sim$ 160 MeV. The mean free path for hadron gas of different sizes approaching the same value at high temperature can be attributed to the cumulative effect of 
reducing differences of the system-size-dependent number density and the system-size independent large interaction cross-sections at high temperature.  \\

\begin{figure}[htb!]
\centering
\includegraphics[scale=0.45]{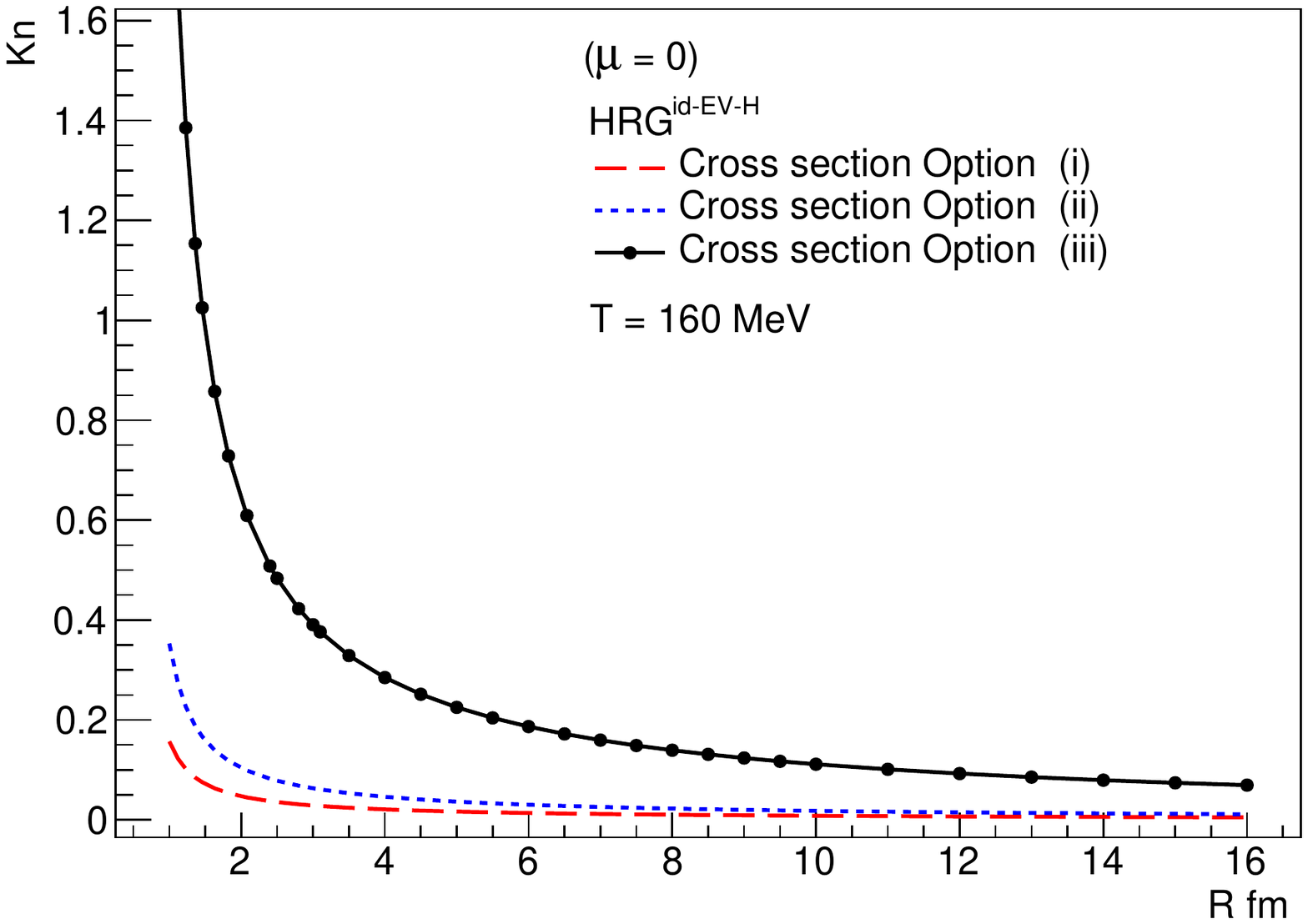}
\caption{The Knudsen number as a function of the system size, R of hadron resonance gas, for different options (described in text) of approximation on cross-sections, 
including the EV effect (and the Hagedorn States for options (i) and (iii)) at $T$ = 160 MeV.} 
\label{fig:kn} 
\end{figure}

By comparing the mean free path and the system-size, one can ideally assess the possibility of thermalization. However, for a quantitative comparison between the 
degree of thermalization of the system formed in high-multiplicity pp events and that of the medium formed in relativistic heavy-ion collisions, we calculate the degree of 
thermalization in terms of the dimensionless Knudsen number, $Kn = {{\lambda} / {R}}$, a ratio between the mean free path and the typical size of the system. A small 
value of $Kn$, tending to zero, implies high degree of thermalization approaching the perfect fluid limit, while a large $Kn$ indicates a system far from thermodynamic 
equilibrium and not suitable for application of fluid dynamics. We calculate $Kn$ as a function of the system size of the considered hadron gas at $T$ = 160 MeV, with 
the ${\lambda}$ calculated from different approximations on $\sigma_{\pi\pi}$, like (i) all hadrons of radius 1 fm leading to the total cross section $\sigma_{T} = \pi$ $fm^{2}$ 
\cite {ref28}, (ii) $\sigma_{meson-meson}$ = 4/9 x ($\sigma_{baryon-baryon}$), $\sigma_{meson-baryon}$ = 2/3 x ($\sigma_{baryon-baryon}$) while $\sigma_{baryon-baryon} 
= \pi$ $fm^{2}$ \cite {ref28} (obviously for the discrete masses only) and iii) pion gas \cite {ref27}, as depicted in Figure~\ref{fig:number}(b). From the  Figure~\ref{fig:kn}, it 
is clear that among the considered approximations, the cross-section option - (iii) gives the most conservative estimates on the system-size dependence of the Knudsen 
number for the hadron gas and we, therefore consider the $Kn$ for this option only for further study / discussions. As can be seen from the Figure~\ref{fig:kn}, there is a 
rapid variation in values of $Kn$ in between the system-size $R \approx$ 4 ($Kn \approx$ 0.28) and 2.5 ($Kn \approx$ 0.48) fm. \\

The Knudsen number for $AuAu$ collisions at $\sqrt s_{NN}$ = 200 GeV have been estimated by hydrodynamic and transport models. In hydrodynamical approach, 
the centrality dependent elliptic flow suggest \cite {ref29, ref30} $Kn \approx$ 0.3 and 0.5 for the central and a semi-central $AuAu$ events, respectively. A good agreement 
between the Boltzmann equation for dilute system and the relativistic dissipative fluid dynamics, the main tool to study the space-time evolution of the bulk matter formed in 
the relativistic heavy-ion collisions at RHIC, is established \cite {ref31, ref32, ref33} for $Kn \textless$ 0.5.\\

To estimate the degree of thermalization in high multiplicity $pp$ events, we compare the sizes of considered hadron gas and the $pp$ events with average charged 
particle multiplicity, $\langle N_{ch} \rangle \approx$ 136 and higher, which exhibit collective properties in terms of the near-side long-range two-particle angular correlations 
\cite {ref05} and also give $p_{T}$-dependence of the elliptic flow harmonic \cite {ref07}, at $\sqrt s$ = 7 TeV. The effective emission radius, $R^{\prime}$, obtained \cite{ref34} 
from the Bose-Einstein Correlations between pairs of identical bosons of the hadronizing system near the kinetic freeze-out, has been parameterized for the $pp$ collisions, 
as a function of $N_{ch}$, as $R^{\prime} (\langle N_{ch} \rangle) = a. \langle N_{ch} \rangle^{1/3}$,  where $a = 0.612 \pm 0.007(stat.) \pm 0.068(syst)$ fm for $pp$ 
collisions at $\sqrt s$ = 7 TeV. So, the high-multiplicity $pp$ event sample of $\langle N_{ch} \rangle$ = 136, corresponding to the average effective emission radius, 
$R^{\prime}$ equals 3.1 fm, may be represented by a system of hadron gas following the LQCD EoS and having the Knudsen number $\sim$ 0.36. These high-multiplicity 
$pp$ events, satisfying the Knudsen number criterion, lie well within the limit of applicability of the fluid dynamics. \\

\section{Summary}
\label{}
In summary, we have studied grand canonical ensemble of hadrons, resonances and Hagedorn states in finite system size, complying with the LQCD equation of state. 
We find, for any finite system-size of HRG that follows the LQCD EoS, the mean free path of pions asymptotically approaches the same  value of 1 fm at $T \approx$ 160 MeV. 
The estimated degree of thermalization of hadron gas of size comparable to the size of high-multiplicity $pp$ events appears consistent with that for the $AuAu$ 
collisions at the RHIC. The study concurs the applicability of hydrodynamics in interpreting the features \cite {ref05, ref06, ref07} of the multiparticle production in 
high-multiplicity $pp$ events at the LHC.\\ 

\section{Acknowledgements}
\label{}
PG acknowledges useful discussions with Jan-s Alam, Partha Pratim Bhaduri and Abhijit Bhattacharyya.\\

\end{document}